\documentclass[twocolumn,showpacs,prc]{revtex4}
\usepackage{dcolumn}
\usepackage{amsmath}
\makeatletter
\providecommand{\LyX}{L\kern-.1667em\lower.25em\hbox{Y}\kern-.125emX\@}

\usepackage[T1]{fontenc}
\usepackage[latin1]{inputenc}
\usepackage{graphics}
\makeatother
\begin{document}

\title{Excitation modes of $^{11}$Li at E$_x \sim$ 1.3 MeV from
proton collisions}

\author{R. Crespo}
\email{raquel@wotan.ist.utl.pt}
\affiliation{Departamento de
F\'{\i}sica, Instituto Superior T\'ecnico, Av.\ Rovisco Pais, 1049-001
 Lisboa, Portugal}
\author{I.J. Thompson}
\email{I.Thompson@surrey.ac.uk}
\affiliation{Department of Physics, University of Surrey,
Guildford,  Surrey, GU2 7XH, United Kingdom}
\author{A.A. Korsheninnikov}
\email{alexei@postman.riken.go.jp }
\affiliation{RIKEN, Wako, Saitama 351-0198, Japan}

\date{\today}

\newcommand{\be}{\begin{eqnarray}}
\newcommand{\ee}{\end{eqnarray}}

\begin{abstract}
The cross section for p-$^{11}$Li inelastic scattering at 68
MeV/u is evaluated using the Multiple Scattering expansion of
the total Transition amplitude (MST) formalism, and compared
with the breakup in the shakeoff approximation. Three different
potential models for $^{11}$Li are used to calculate the
$^{11}$Li(p,p$'$) continuum excitations, and all show peaks
below 3 MeV of excitation energy,
both in resonant and some non-resonant channels.
In the most realistic model of $^{11}$Li, there is a strong dipole
contribution associated with attractive but not a fully-fledged resonant phase
shifts, and some evidence for a  J$_{nn}^\pi = 0_2^+$ resonant
contribution. These together form a
pronounced peak at around 1--2 MeV excitation, in agreement
with experiment, and this supports the use of the MST as an adequate formalism
to study excited modes of two-neutron nuclear halos.

\end{abstract}

\maketitle

Halo nuclei are weakly bound structures in a vicinity of a breakup
threshold, and the knowledge of the continuum properties
is an essential tool for the understanding of these nuclei.
These  structures are of wide interest in other fields such
atomic and molecular as well as nuclear physics 
\cite[and references therein]{Hansen01}.

One timely issue is whether the correlations between
the cluster systems of the halo nuclei are sufficiently strong
to support excited states or resonances in the continuum.
In particular, it is still an open problem whether there exists
a new kind of collective motion,
the `soft dipole' excited state or resonance at low energies in the
breakup continuum, as predicted by some theories \cite{soft}.
Some evidence for these modes was found for the
Borromean two-neutron halo nucleus $^{11}$Li
\cite{Alexei,soft11Li} and $^6$He \cite{nakayama}, but an unequivocal
signature remains to be found.
A detailed study of the resonances  in the continuum sea, which has just
now begun to be possible, would help shed
light on the existence of the excited modes of halo nuclei,
and on other related issues.
The study of these modes is also relevant for the comprehension
of the  ground state structure, because the mechanisms
for the halo excitation depend on the ground state properties.

The aim of this work to study the evidence of low lying excited states
in $^{11}$Li in inelastic collisions from protons within the few-body Multiple
Scattering expansion of the total Transition amplitude (MST)
formalism \cite{MST} using different few-body potential models
for the $^{11}$Li ground state and continua, and results compared with
those of simpler breakup models.

Several structure models have been developed to describe the structure
of $^{11}$Li \cite{thom94,thom98,Cobis,Karataglidis}.
These  calculations predict different resonances for the
valence neutron halo pair. In particular, it is unclear if the
neutron-neutron and neutron-$^9$Li correlations are together
sufficiently strong to constitute a soft dipole resonance $J_{nn}^\pi = 1^-$.
Moreover, a low lying resonance $J_{nn}^\pi = 0_2^+$  was predicted
in \cite{thom98,Cobis}, but no evidence for this has been found up to now.

In parallel to the theoretical analyses, the low lying excited states
of $^{11}$Li have been experimentally investigated
\cite{Alexei,Kobo,Gornov,Bohlen}.
These very difficult studies, suffering in some cases from poor
statistics, have shown contradictory results, in particular
with respect to the existence of a low lying excited state
at E$^* \sim$ 1.3 MeV. Inelastic scattering from protons  \cite{Alexei}
can be a tool to find evidence for low lying states.
The evident interplay between the extracted structure information
and the scattering approach \cite{Alexei,Karataglidis}
calls for a clarification of the scattering
framework when describing the scattering from halo nuclei.

Traditional calculations of inelastic cross sections assume
collective excitations and use optical model potentials, with
few-body dynamics perhaps only included approximately by means of
effective interactions.
The halo degrees of freedom can be explicitly incorporated
in the scattering framework in a convenient way within
the MST approach \cite{MST,clus,evora}, and this has the advantages of
including couplings
to the continuum in all orders, of clearly delineating the
structure and dynamics, and of treating up to four body problems \cite{MST}.
Alternative coupled channel approaches \cite{CDCC}, which explicitly expand
on continuum states, are only able to tackle up to three body problems.

We consider then the scattering of a nucleon (particle 1) from  $\cal N$
projectile subsystems. In the case of $^{11}$Li,
assumed to be well described by a three-body
($^9$Li + $n$ + $n$) model, $\cal N$=3.
The total transition amplitude
$T$ can be written as a multiple scattering expansion in the transition
amplitudes $\hat{t}_{\cal I}$ for proton
scattering from each projectile sub-system ${\cal I}$ \cite{MST}
\begin{equation}
T = \sum_{\cal I}\hat{t}_{1{\cal I}}  +
 \sum_{\cal I} \hat{t}_{1 {\cal I}} G_0 \sum_{{\cal J} \neq {\cal I}}
  \hat{t}_{1 {\cal J}} + \cdots
\label{TMSexp}
\end{equation}
where the propagator $G_0 =\left( E^+ - K \right)^{-1}$, within
 the impulse approximation, contains the
kinetic energy operators of the projectile and all
the target subsystems.
Here $E$ is the kinetic energy in the
overall center of mass frame \cite{MST}.
It follows from Eq. (\ref{TMSexp}) that
in the MST expansion the few-body dynamics is properly included, and
excitations of the  projectile which involve changes in the relative motion
of the sub-systems are taken into account.
The contribution of these to the calculated elastic cross
section was investigated in \cite{MST,clus,adiabatic}.

We have in mind the scattering process of  $^{11}$Li, originally in a
$|\phi_0 \rangle$ state, to a final $|\phi_f \rangle$ state, by means of its
interaction with a proton, with initial momentum $\vec{k}_i$  and final
momentum $\vec{k}_f$ in the nucleon-nucleus  center-of-mass frame.
We describe the final state with angular  momentum of the
valence neutron pair $J_{nn}^\pi(f)$ and excitation energy $E^*_f$ as
$|\phi_f \rangle = |J_{nn}^\pi(f), E^*_f \rangle$, neglecting
the spin of the core.
In the experiment reported by \cite{Alexei},
final states  up to an excitation energy $E^*_f \leq 15 $ MeV were detected.
Previous studies \cite{thom98} of $^{11}$Li excitations show that it is sufficient
to include contributions from dipole $J^\pi_{nn} = 1^-$,
 spin-dipole  $J^\pi_{nn} = 0^-$, spinflip
$J^\pi_{nn} = 1^+$, and second $J^\pi_{nn} = 0_2^+$,
excitations in the scattering.

We use a single scattering approximation, so Eq. (\ref{TMSexp}) reduces to:
\be
T =  \hat{t}_{1 {\rm core}} +  \sum_{n=1,2}\hat{t}_{1 n}~~,
\label{tall}
\ee
where $\hat{t}_{1 {\rm core}}$, $\hat{t}_{1 n}$ are the transition amplitudes
for the scattering from the core and valence neutrons respectively.

In the work of Karataglidis {\em et al} \cite{Karataglidis}
the differential cross section is calculated using the shakeoff
approximation (SA).
This consists first of all, in taking into account
only the proton-core contribution to the single
scattering term, so Eq. (\ref{TMSexp}) becomes
\begin{equation}
\langle \vec{k}_f \phi_f | T | \phi_0 \vec{k}_i \rangle =
 \hat{t}_{1 {\rm core}}(\omega,q)  F_{f0}(\alpha q) \label{SA}
\end{equation}
where $\vec{q} =  \vec{k}_f -  \vec{k}_i$. In this equation,
$F_{f0}(\alpha q)$ is the transition  density,
and $\alpha= 2/11$ \cite{MST}.
The transition amplitude $ \hat{t}_{1 {\rm core}}(\omega,q)$ describes the
scattering from the $^9$Li core at the appropriate energy $\omega$.
Then, summing the contributions
of {\em all} the continuum for the scattering process and using closure,
the inelastic cross section is
\begin{equation}
\left( \frac{d\sigma}{d\Omega} \right)  _{\rm SA} = {\cal R}
\left( \frac{d\sigma}{d\Omega}\right)_{9}(1-|F_{00}(\alpha q)|^2)~~,
\label{shake}
\end{equation}
where $\left( {d\sigma}/{d\Omega}\right)_{9}$ is the differential elastic
cross section for p-$^9$Li scattering and $F_{00}(\alpha q)$ the density
distribution for the motion of the core centre-of-mass \cite{MST}.
The departure of $F_{00}(\alpha q)$ from unity
at nonzero transferred momentum $q$
arises from core recoil effects \cite{MST,adiabatic}.
The renormalization factor ${\cal R}$ is somewhat arbitrary
and is chosen to remove all the contributions from the continuum
that are excluded by the experimental acceptance.

We show the results of  MST calculations that take into account
the contributions where the proton scatters both from the core
and from the valence neutrons in Eq. (\ref{tall}), and we do not use
the closure approximation.

\begin{figure}
{\par\centering \resizebox*{0.40\textwidth}{!}
{\includegraphics{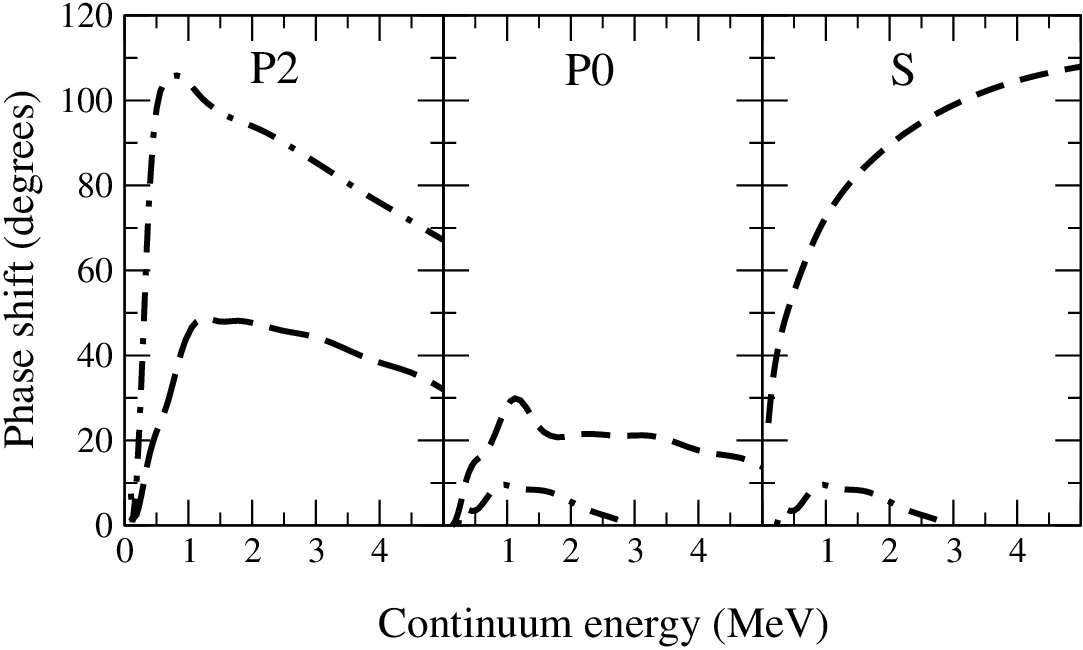}} \par}
\caption{\label{Fig:fig1}
Calculated phase shifts for the 3  structure model.
The dashed line represents the $K$=1, $S$=0 channel for J$_{nn}^\pi = 1^-$
the dashed-dotted the $K$=0, $S$=0 channel for  J$_{nn}^\pi = 0_2^+$,
for three-body hypermoment $K$ and two-neutron spin $S$
as in  \protect\cite{dan98,thom98}. }
\end{figure}

In describing $^{11}$Li, the internal
and spin dynamical properties of the $^9$Li core are included  approximately
through a nucleon-core effective interaction, and then
the ground state and continuum wave functions are obtained by solving the Faddeev equations.
We consider here three structure models for which all eigenstates
are defined by different sets of n-core potentials. All models
use the GPT nn potential \cite{GPT}. The first model (S) uses n-core potentials
from Johannsen, Jensen and  Hansen \cite{jjh}, and gives an
$s^2$-dominated $^{11}$Li wave function
similar to that used in the shakeoff calculations of \cite{Karataglidis}.
The second and third models are defined in \cite{thom94}, and include
Pauli blocking operators for the $s_{1/2}$ and $p_{3/2}$ core states.
The second model (P0) uses potentials similar to those of Bertsch and Esbensen
\cite{be91},
and gives $(0p_{1/2})^2$ halo wave functions as would be expected from normal shell model
ordering. A final model (P2) is that advocated by Thompson and Zhukov \cite{thom94},
having an $s$-wave mixture arising from $sd$  intruder levels in $^{10}$Li.
The intruder levels have a profound effect on the $^{11}$Li structure
\cite{thom94,thom98}, and the P2 model contains a superposition of
$(0p_{1/2})^2$ and $(1s_{1/2})^2$ components with relative
weights of 45$\%$ and 31$\%$, in good agreement with \cite{simon}.

The dominant hyperspherical phase shifts are shown in Fig. (\ref{Fig:fig1}),
calculated using the methods of \cite{dan98}.
According to \cite{thom98} and these calculations, within the P2 model
a low lying resonance can be found at $E^*_f = 0.5 $ MeV
of width $\Gamma = 0.6 $ MeV for $J_{nn}^\pi = 0_2^+$,
as a superposition of $s^2$ and $p^2$
configurations orthogonal to those of the ground state,
and such a resonance is not predicted in the case of the P0 and S models.
There are enhanced soft-dipole $1^-$
final state interactions in both the P2 and S models,
from $s_{1/2}p_{1/2}$ neutron states.
In this channel the phase shift does rise rapidly, but only 
shows at most a resonant-like behaviour,
and strictly there is no dipole resonance in any of the models.
In all three models, however, a large non-resonant contribution is
expected that arises primarily from the large size of the $^{11}$Li
ground state, but the size of any such transitions will be enhanced
by positive continuum phase shifts.

\begin{figure}
{\par\centering \resizebox*{0.38 \textwidth}{!}
{\includegraphics{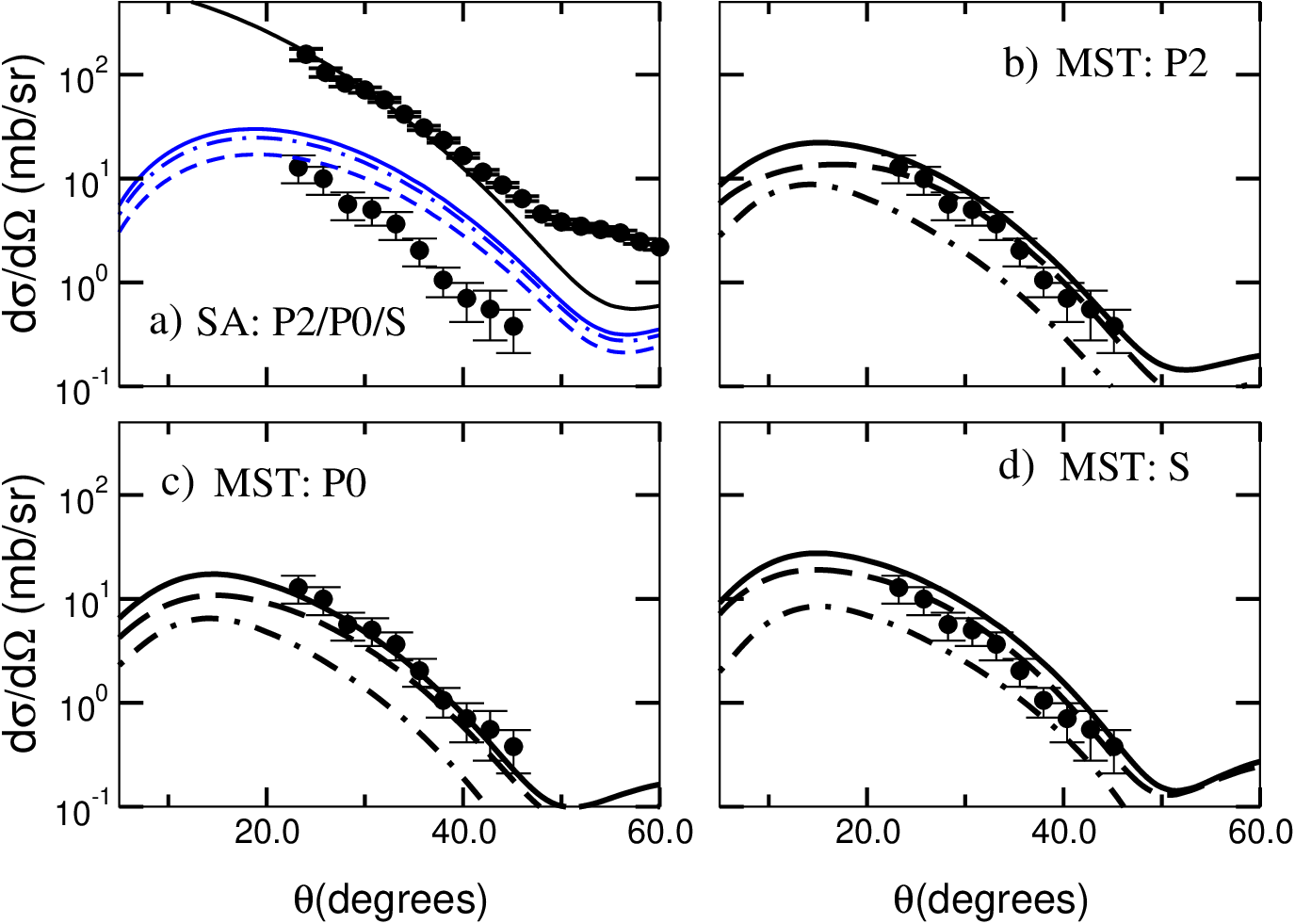}} \par}
\caption{\label{Fig:figxs-models.mcarlo}
Calculated inelastic cross section for p-$^{11}$Li
inelastic cross section at 68 MeV/u within the MST framework
using the models for $^{11}$Li described in the text.}
\end{figure}

The transition amplitude for proton scattering from the $^{9}$Li core
was generated by the multiple scattering expansion of the  optical potential
in terms of the free NN transition amplitude, calculated in the
single scattering approximation \cite{Li11} with only a central interaction
and neglecting the Coulomb interaction since this is only relevant
at very low angles.
We use the on-shell approximation for the matrix elements of the
transition amplitude in  momentum space,
which should be
a reasonable approximation in this energy regime and for low excitation
energies.
In the  evaluation of the contribution from the valence neutrons, the
spin dependence of the NN  amplitudes given by
the tensor representation of \cite{tensor}.

The $^{9}$Li ground state was taken as in
 \cite{evora}, which provides a reasonable
description of the p-$^9$Li elastic data \cite{Moon} in the angular region
$\theta \leq 40 ^\circ$
as shown in the upper curve of Fig.\ (\ref{Fig:figxs-models.mcarlo}a).

The experimental differential cross sections from \cite{Alexei} for
p-$^{11}$Li inelastic
scattering at 68 MeV/u are shown in Fig.\ (\ref{Fig:figxs-models.mcarlo}).
We also show in Fig.\ (\ref{Fig:figxs-models.mcarlo}a) the inelastic scattering within the
shakeoff approximation (SA), Eq.\ (\ref{shake}).
The results for the three models with ${\cal R}=1$ are represented
by the  solid (P2) the  dashed (P0) and the dashed-dotted (S) lines,
and are all more than twice the experimental magnitude
in the region where the scattering from the core is well described.
Even when introducing a renormalization factor, ${\cal R}$, as
in \cite{Karataglidis}, the calculated cross sections using the
shakeoff framework decay more slowly than
the  data, and thus do not give a good description
of the scattering.

The calculated inelastic angular distributions using MST with the three
structure models are plotted in
Figs.\ (\ref{Fig:figxs-models.mcarlo}b-d),
by integrating  $d^2\sigma/d\Omega dE_f$
over the experimentally defined section of the energy spectrum \cite{Alexei},
where we have  calculated all the excited (resonant and non-resonant)
contributions.
In these figures, the solid line includes the sum of the $J^\pi_{nn} = 1^-$
and   $J^\pi_{nn} = 0_2^+$  contributions. The inelastic transitions calculated
with  $1^-$ and  $0_2^+$ only are represented
by dashed  and dashed-dotted lines  respectively.
The contributions from the  spin-dipole  $0^-$ and spinflip $1^+$
excited  states do not significantly alter the full calculations,
and are thus not shown in  Fig. (\ref{Fig:figxs-models.mcarlo})
for simplification.
When comparing the dashed and solid lines, it is evident that the
major contribution comes from the
dipole mechanism, with a small contribution from excited $0_2^+$ states.
For all three structure models the total differential cross section using the
MST scattering framework agrees well with the available data.

\begin{figure}
{\par\centering \resizebox*{0.38\textwidth}{!}
{\includegraphics{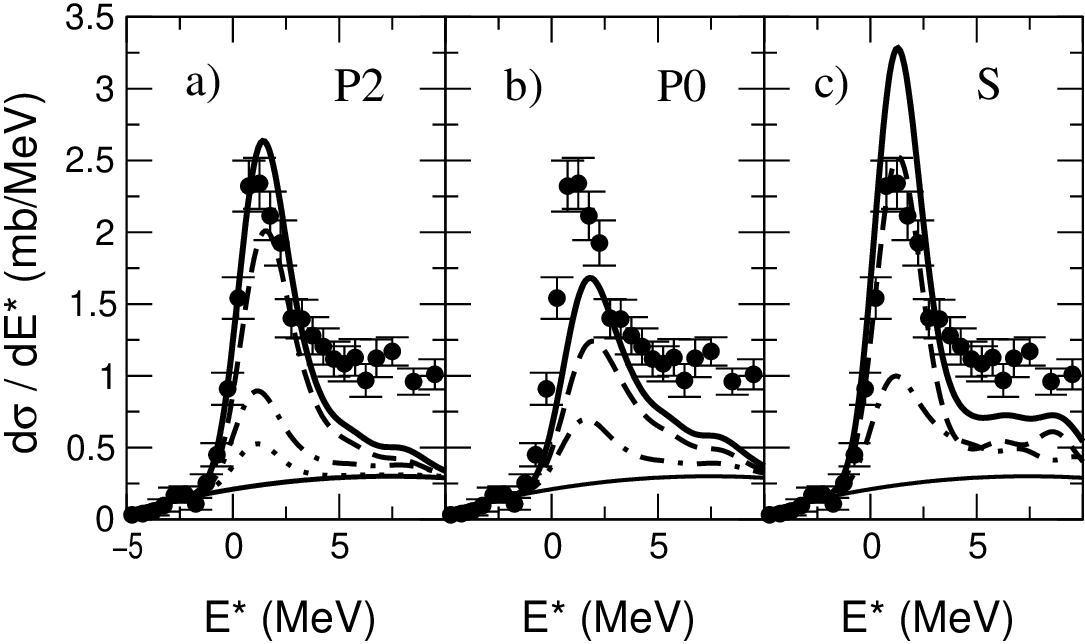}} \par}
\caption{\label{Fig:figexc-models.mcarlo}
Calculated energy spectrum for the 3 models described in the text.
The dashed line represents the J$_{nn}^\pi = 1^-$
the dashed-dotted the J$_{nn}^\pi = 0_2^+$ and the  solid line the sum.
The lower thin solid line shows the background from materials other
than protons in the target. In a) the dotted curve gives
the calculated spectrum with no contribution from the scattering of the 
valence neutrons to the J$_{nn}^\pi = 0_2^+$ state.
 }
\end{figure}

We now analyse the energy spectrum, in Fig. (\ref{Fig:figexc-models.mcarlo}).
The double differential cross section $d^2\sigma/d\Omega dE_f$
was calculated up to 10 MeV; angular  acceptance and energy resolution of the
proton detection system was incorporated by simulation
of the experimental apparatus. The experimental numbers of counts given in
\cite{Alexei} are here converted to cross sections in mb/MeV.
In Fig.\ (\ref{Fig:figexc-models.mcarlo}), as in the case of the
Figs.\ (\ref{Fig:figxs-models.mcarlo}b-d)
the dashed line includes only the dipole contribution
and the dashed-dotted the 0$^+_2$ contribution. The sum is given
by the solid line. The other states give a small contribution to
the energy spectrum, and therefore are not included.
In the experiment, CH$_2$ was used as a target, and the thin solid line
in Figs.\ (\ref{Fig:figxs-models.mcarlo}) shows the background from materials
other than protons in the target.

All of the models fail to reproduce the cross sections above
$\sim$ 5 MeV, indicating that further
mechanisms are occurring that are outside the scope of our
few-body model, or that higher order terms of the multiple
scattering expansion might have been important.
However, the peak below 5 MeV {\em can} be
reproduced, to varying degrees of accuracy in the various
models, indicating that some structure information
can be extracted from the very precise low energy spectrum.
In the case of the P2 model
Fig. (\ref{Fig:figexc-models.mcarlo}a), the dipole contribution underestimates
the energy spectrum. When including the second 0$_2^+$ represented
by the dashed dotted line, however, the total spectrum with the
two contributions  reproduces well
the low energy data. As for the P0 model
Fig. (\ref{Fig:figexc-models.mcarlo}b), even when including the
0$_2^+$ contribution, the predicted energy spectrum underestimates the data.
On the other hand, when including this state the S model
Fig. (\ref{Fig:figexc-models.mcarlo}c) overestimates the experimental points.
In Fig. (\ref{Fig:figexc-models.mcarlo}a)
the calculated spectrum with no contribution from the scattering from
the valence neutrons to the
second 0$_2^+$ resonant state is represented by the dotted curve.
The difference between this and  the  dashed-dotted curve shows that
the scattering from the valence nucleons is essential for the
0$_2^+$ excitation.
This contribution was not taken into account in the
SA framework, in order to permit the closure summation.

We conclude that the shakeoff framework fails to describe both the shape
and magnitude of the inelastic cross section, and find
that MST is a useful scattering framework to obtain information about
halo excitation modes from accurate inelastic energy spectrum data.

When considering the low lying energy spectrum
up to 5 MeV,  the P2 structure model
for $^{11}$Li  reproduces well the differential cross section and
the shape, position and magnitude of the peak.

We see that the experimental data can be well
reproduced by a
three-body model of $^{11}$Li in which there is a pronounced
$1^-$ peak at low continuum energies, but yet in which there is not
a fully-fledged resonance in this breakup channel.
There is a
0$_2^+$ resonance which contributes to this peak, but most of
the cross section arises from the $1^-$ nuclear dipole
excitation mechanism.
This is in partial agreement with \cite{Karataglidis},
though here we do see definitive effects of attractive
final-state interactions, as reflected in the continuum phase shifts
of a more realistic $^{11}$Li model (dashed curve in Fig. 1 for the P2 model).

Experimental evidence for the existence of a strong (but
not a fully-fledged resonant) dipole peak is a further demonstration of the novel
range of phenomena that occur already with three bodies
in quantum few-body dynamics. Furthermore, this work shows some
first evidence of a 0$_2^+$ resonance contribution at 1-2 MeV excitation.

\bigskip \bigskip

{\bf Acknowledgements:}
Work supported by
Funda\c c\~ao para a Ci\^encia e Tecnologia
(Portugal) through grant No. POCTI/1999/FIS/36282 and
FMRH/BSAB/125/99,
and in the U.K. by EPSRC grant GR/M/82141.


\end{document}